\documentclass[english,a4paper,color]{cg}
\pdfoutput=1
\usepackage{babel}
\usepackage{mathptmx}
\usepackage{graphicx}
\usepackage{amssymb}
\usepackage{amsmath}
\usepackage{a4wide}
\usepackage{color}
\usepackage{fancyheadings}
\usepackage{paralist}
\usepackage[utf8]{inputenc}
\usepackage[T1]{fontenc}
\usepackage{hyperref}

\hypersetup{
pdftitle={User Interface for Volume Rendering in Virtual Reality Environments},
pdfsubject={Technical Report},
pdfauthor={Jonathan Klein, Dennis Reuling, Jan Grimm, Andreas Pfau, Damien Lefloch, Martin Lambers, Andreas Kolb},
colorlinks=true,
linkcolor=black,
citecolor=black,
filecolor=black,
urlcolor=black,
}

\renewcommand{\thesection}{\arabic{section}}
\renewcommand{\thesubsection}{\thesection.\arabic{subsection}}
\renewcommand{\thesubsubsection}{\thesubsection.\arabic{subsubsection}}
\renewcommand{\theparagraph}{\thesubsubsection.\arabic{paragraph}}

\makeatletter
\def\fnum@table{Table \thetable}
\def\fnum@figure{Figure \thefigure}
\renewcommand\section{\@startsection {section}{1}{\z@}%
                                   {-3.5ex \@plus -1ex \@minus -.2ex}%
                                   {2.3ex \@plus.2ex}%
                                   {\normalfont\Large\bfseries}}
\renewcommand\subsection{\@startsection{subsection}{2}{\z@}%
                                     {-3.25ex\@plus -1ex \@minus -.2ex}%
                                     {1.5ex \@plus .2ex}%
                                     {\normalfont\large\bfseries}}
\renewcommand\subsubsection{\@startsection{subsubsection}{3}{\z@}%
                                     {-3.25ex\@plus -1ex \@minus -.2ex}%
                                     {1.5ex \@plus .2ex}%
                                     {\normalfont\normalsize\bfseries}}
\renewcommand\paragraph{\@startsection{paragraph}{4}{\z@}%
                                    {3.25ex \@plus1ex \@minus.2ex}%
                                    {-1em}%
                                    {\normalfont\normalsize\bfseries}}
\renewcommand\subparagraph{\@startsection{subparagraph}{5}{\parindent}%
                                       {3.25ex \@plus1ex \@minus .2ex}%
                                       {-1em}%
                                      {\normalfont\normalsize\bfseries}}
\makeatother

\begin{document}

\dbDocumentType{~}
\dbTitle{User Interface for Volume Rendering\\in Virtual Reality Environments}
\dbAuthor{Jonathan Klein, Dennis Reuling, Jan Grimm, Andreas Pfau,\\
Damien Lefloch, Martin Lambers, Andreas Kolb}
\dbVersion{February 8, 2013}
\dbDisclosure{0}

\cgDeckblatt

\setlength\headsep{.275in}
\newpage

\section*{Abstract}
Volume Rendering applications require sophisticated user interaction for the
definition and refinement of transfer functions. Traditional 2D desktop user
interface elements have been developed to solve this task, but such concepts do
not map well to the interaction devices available in Virtual Reality
environments.

In this paper, we propose an intuitive user interface for Volume Rendering
specifically designed for Virtual Reality environments. The proposed interface
allows transfer function design and refinement based on intuitive two-handed
operation of Wand-like controllers. Additional interaction modes such as
navigation and clip plane manipulation are supported as well.

The system is implemented using the Sony PlayStation Move controller system.
This choice is based on controller device capabilities as well as application and
environment constraints.

Initial results document the potential of our approach.

\section{Introduction}

Volume Rendering visualizes 3D grids of voxels. Each voxel typically stores a
scalar value representing density, as retrieved via a 3D scanning technique such
as CT or MRI.
Direct Volume Rendering techniques such as volume ray
casting work directly on the voxel data instead of extracted geometry such as
isosurfaces. Such techniques use a \emph{transfer function} to map voxel values
to opacity and color. The volume ray caster then generates a ray through the 3D
grid for every pixel in the image plane, samples the voxel data along the ray,
and composites the opacity and color information given by the transfer function
to compute the final pixel color.

A basic transfer function is a one-dimensional function that directly maps a
scalar voxel value to opacity and color.
Volume Rendering applications require user interface concepts that allow
efficient and precise design and refinement of such transfer functions, to
enable the user to visualize the interesting parts of the volume data set.
In the traditional 2D graphical user interface domain of desktop systems, this
problem is solved using 2D widgets that typically allow mouse-based manipulation
of the functions~\cite{engel06volume}.
This paper focuses on one-dimensional transfer functions, but note that advanced
two-dimensional transfer functions models exist that take the gradient or the
curvature at the voxel location into account and require even more complex user
interfaces.

Since Virtual Environments are especially well suited to explore spatial
properties of complex 3D data, bringing Volume Rendering applications into such
environments is a natural step. However, defining new user interfaces suitable
both for the Virtual Environment and for the Volume Rendering application is
difficult. Previous approaches mainly focused on porting traditional 2D
point-and-click concepts to the Virtual
Environment~\cite{schulze01volrend,kniss04medapp,shen08medvis}. This tends to be
unintuitive, to complicate the interaction, and to make only limited use of
available interaction devices.

In this paper, we propose an intuitive 3D user interface for Volume Rendering
based on interaction devices that are suitable for Virtual Reality environments. We focus on 
a simplified approach to design and refine transfer functions that allows
intuitive use of interaction devices, specifically the Sony PlayStation Move
controller system. Our demonstration system also supports other Volume Rendering
interaction modes such as navigation and clip plane manipulation.

The remainder of this paper is organized as follows. Sec.~\ref{sec:related-work}
discusses related work. In Sec.~\ref{sec:concept}, we describe our user
interface concepts in detail, and present its implementation based on Sony
PlayStation Move controllers in a Virtual Reality Lab. Initial results are shown
in Sec.~\ref{sec:results}.  Sec.~\ref{sec:conclusion} concludes this paper.

\section{Related Work}
\label{sec:related-work}

One of the first applications of Volume Rendering in a Virtual Reality
environment was presented by Brady et al.~in 1995~\cite{brady95crumbs}. This
early work concentrated on navigation using a Wand-like device.
In 2000, Wohlfahrter et al.~presented a two-handed interaction system with
support for navigating, dragging, scaling, and cutting volume data in a Virtual
Reality environment~\cite{wohlfahrter00studydesk}.
Neither of these early approaches supported transfer function manipulation.

One of the first works on transfer function editing for Volume Rendering in
Virtual Reality environments was presented by Schulze-Döbold et
al.~in 2001~\cite{schulze01volrend}.  Their transfer function editor requires a
6 DOF controller with three buttons. The controller is primarily used to point
at an interaction element to select it for manipulation. To control scalar
values, the editor uses virtual knobs that are manipulated by twisting the hand. The
three buttons are used to manipulate position and size of the 2D transfer
function editor inside the 3D environment. This interface is directly based on
the 2D desktop point-an-click interface. Consequently, the authors refer to the
controller as a 3D mouse.  Schulze-Döbold later refined the user interface based
on feedback collected in a user study~\cite{schulze03phd}, but the principal design
remained unchanged.

Wössner et al.~reuse Schulze-Döbold's work for the purpose of collaborative
volume exploration in distributed setups~\cite{woessner02collab}.
Kniss et al.~split the task of defining multidimensional transfer functions into a
classification step and an exploration step~\cite{kniss04medapp}. The
classification step, which defines the transfer function, is performed prior to
visualization on a classical 2D desktop system using the mouse. The Virtual
Reality interface is based on Schulze-Döbold's work.

Later works also mainly use variations of this approach of bringing 2D
point-and-click interfaces to 3D environments~\cite{he07hci,shen08medvis}.
An exception is the work of Tawara and Ono from 2010, in which they combined a
Wiimote and a motion tracking cube to get a tracked manipulation device for a
volume data application~\cite{tawara10volseg}. However, their approach focuses
on volume segmentation in augmented reality; in particular, it does not support
transfer function manipulation.

\section{3D User Interface for Volume Rendering}
\label{sec:concept}

A user interface for Volume Rendering applications must support two key
interaction modes:
\begin{compactitem}
\item \emph{Navigation.} This allows to inspect the volume from various
perspectives by applying translations and rotations. A Virtual Reality
environment with user tracking additionally allows the user to move around
the volume.
\item \emph{Transfer function manipulation.} A transfer function allows to
visualize the interesting parts of the volume (by assigning color and opacity to
the interesting voxel value ranges) and at the same time remove other parts of
the volume from view (by mapping the corresponding voxel values to zero
opacity).
\end{compactitem}
In addition to these modes, Volume Rendering applications usually provide
supplemental tools such as clip planes.

\begin{figure}[t]
	\centering
	\includegraphics[width=.66\linewidth]{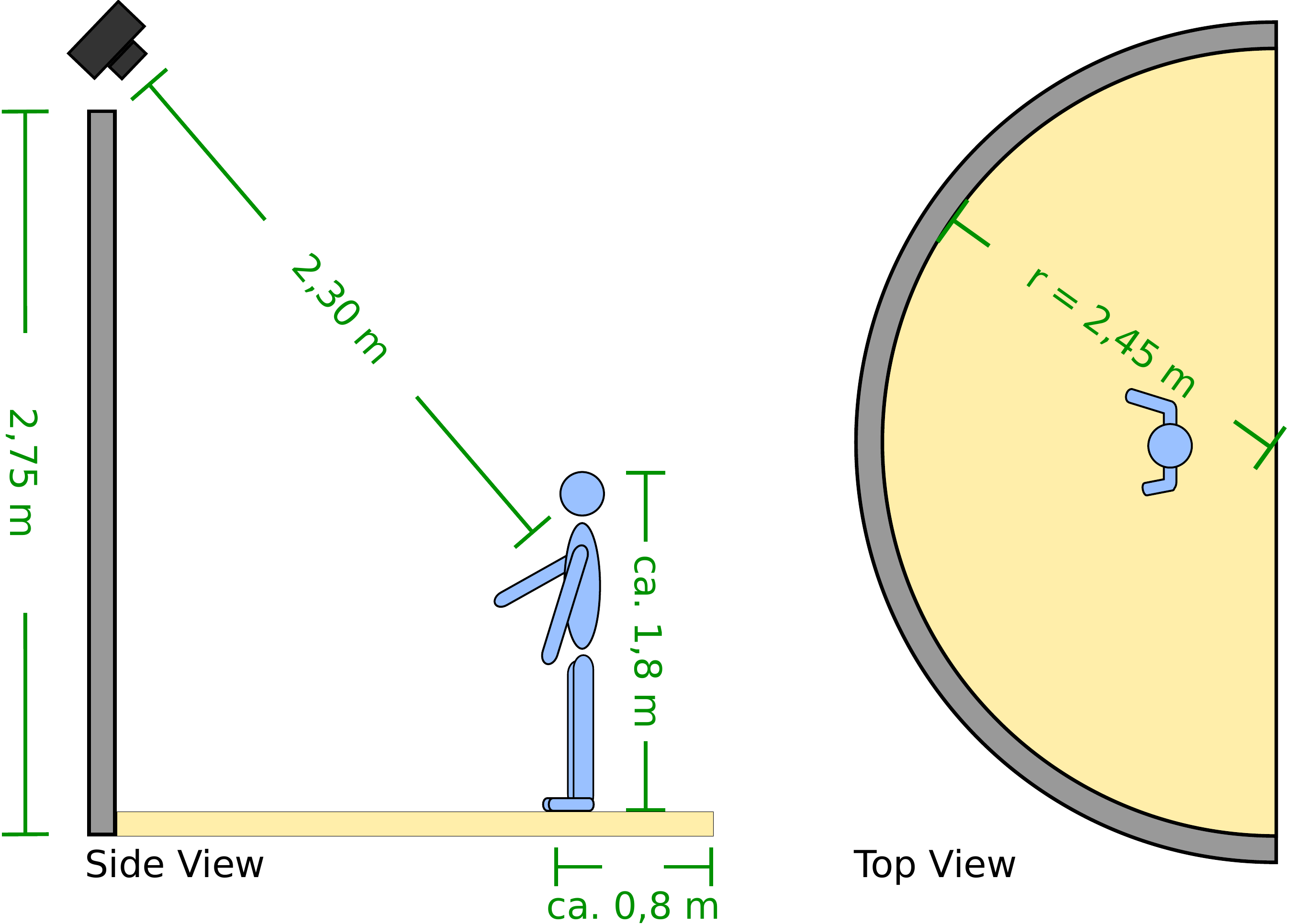}
	\caption{Overview of the Virtual Reality environment}
	\label{vrlab_constraints}
\end{figure}

In this paper, we focus on the transfer function manipulation mode of the user
interface.

\subsection{Choice of Input Devices}

Our choice of input devices, interface concepts, and implementation details was
based on the interaction requirements given by the Volume Rendering application,
with special emphasis on transfer function manipulation, and on the
specific constraints given by the Virtual Reality environment.

\begin{figure}[t]
	\centering
	\includegraphics[width=.66\linewidth]{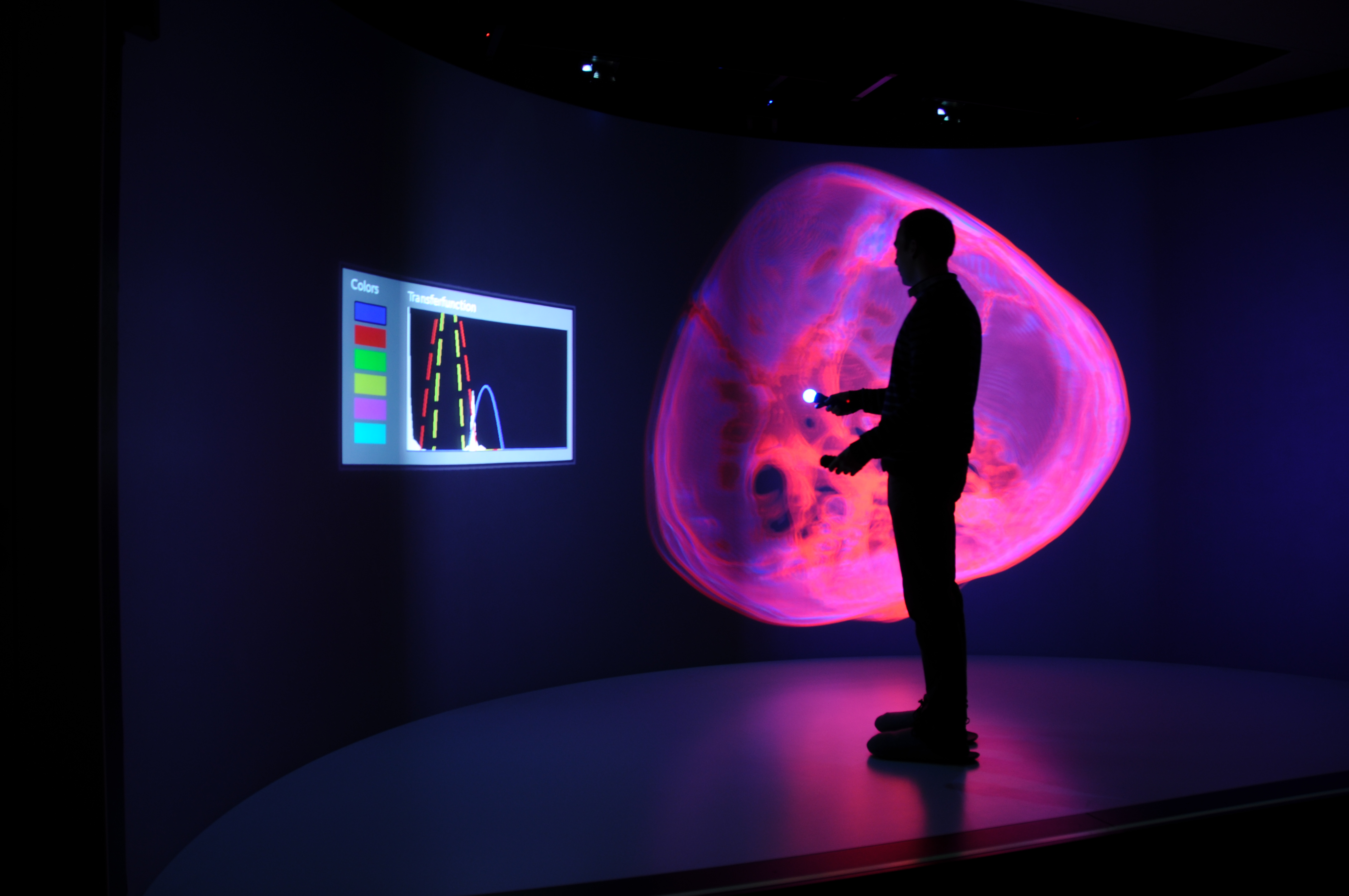}
	\caption{Interactive Volume Rendering in the Virtual Reality environment}
	\label{vrlab_interactive}
\end{figure}

The Virtual Reality environment which we used in this project has an open
cylindrical screen and a floor screen, providing a wide field of view. See
Fig.~\ref{vrlab_constraints} and~\ref{vrlab_interactive}. The cylindrical screen
has four rear-projection stereo channels and the floor screen has two
front-projection stereo channels. Additionally, the environment provides optical user
tracking.

We considered interaction devices that are based on readily available and
affordable hardware components and provide enough input facilities for both
navigation and transfer function editing. Devices that fulfill these criteria
include the Microsoft Kinect, the Nintendo Wiimote, and the Sony PlayStation Move.
Traditional game controllers as well as the Flystick provided by our optical
tracking system were excluded since they do not provide enough input facilities.

Microsoft Kinect uses an optical tracking system to follow the movement of
the user, allowing full-body gesture interaction. Unfortunately the Kinect
system requires the camera to be placed directly in front of the user, which was
not possible in our environment. We experimented with a Kinect camera placed at
the top of the screen and looking at the user with an angle of approximately 45
degrees, but this setup leads to unusable tracking data.

The concepts of the Wiimote and the Move are similar. In both systems the user
holds a wand-like controller in his hand that can measure its movement and
orientation. The Move system additionally uses an optical tracking system to
determine the absolute position of the controller, and can therefore provide
position and orientation data with higher precision and reliability.
In contrast to the Kinect system, the Move system works fine with the camera
mounted on top of the screen, as shown in Fig.~\ref{vrlab_constraints}.
Furthermore, the Move controller has a glowing bulb on top whose color can be
changed. This gives interesting opportunities for user feedback (see
Sec.~\ref{sec:tf-editing}).

Recently, Takala et al.~identified a lack of user interface concepts based on
the PlayStation Move system, partly due to the locked-up nature of the
SDK~\cite{takala12survey}. That situation has changed: the SDK is now freely
available for educational and research purposes.
Additionally, prices for the hardware part of system dropped significantly.

For these reasons, we chose to base our experiments on the PlayStation Move system.
 
\subsection{Interface Concept}

To allow both navigation and transfer function manipulation and to have enough
input facilities for the various functionalities required by these two modes, we
decided to use both the Move controller (for the right hand) and an additional
Move Nav-Pad (for the left hand). See Fig.~\ref{fig:move}.
The Move controller, whose bulb is tracked by the camera of the system, is
used for manipulation (translation and rotation in navigation mode, and
modification in transfer function editing mode), while the Move Nav-Pad is
used for selection and switching purposes.
This configuration is classified by Schultheis et~al.~\cite{schultheis12twohand}
as an asymmetric two-handed interface using two Wand-like input devices.

\subsubsection{Transfer Function Editing}
\label{sec:tf-editing}

Small changes to a transfer function can result in significant changes in the
visualization result, and usually more than one range of voxel values represents
interesting parts of the volume data. For this reason, typical Volume Rendering
applications allow to specify the transfer function using piecewise linear
building blocks. However, this requires a very fine-grained control, typically
using a Mouse-based interface.

In order to allow more intuitive manipulation of transfer functions using
hand-held devices that typically favor coarser control movements, we use a
different transfer function model. In our model, a transfer function is defined
as the sum of window functions, called \emph{peaks}. Each peak highlights a
small voxel value range in a specific color.

\begin{figure}[t]
	\centering
	\includegraphics[width=.66\linewidth]{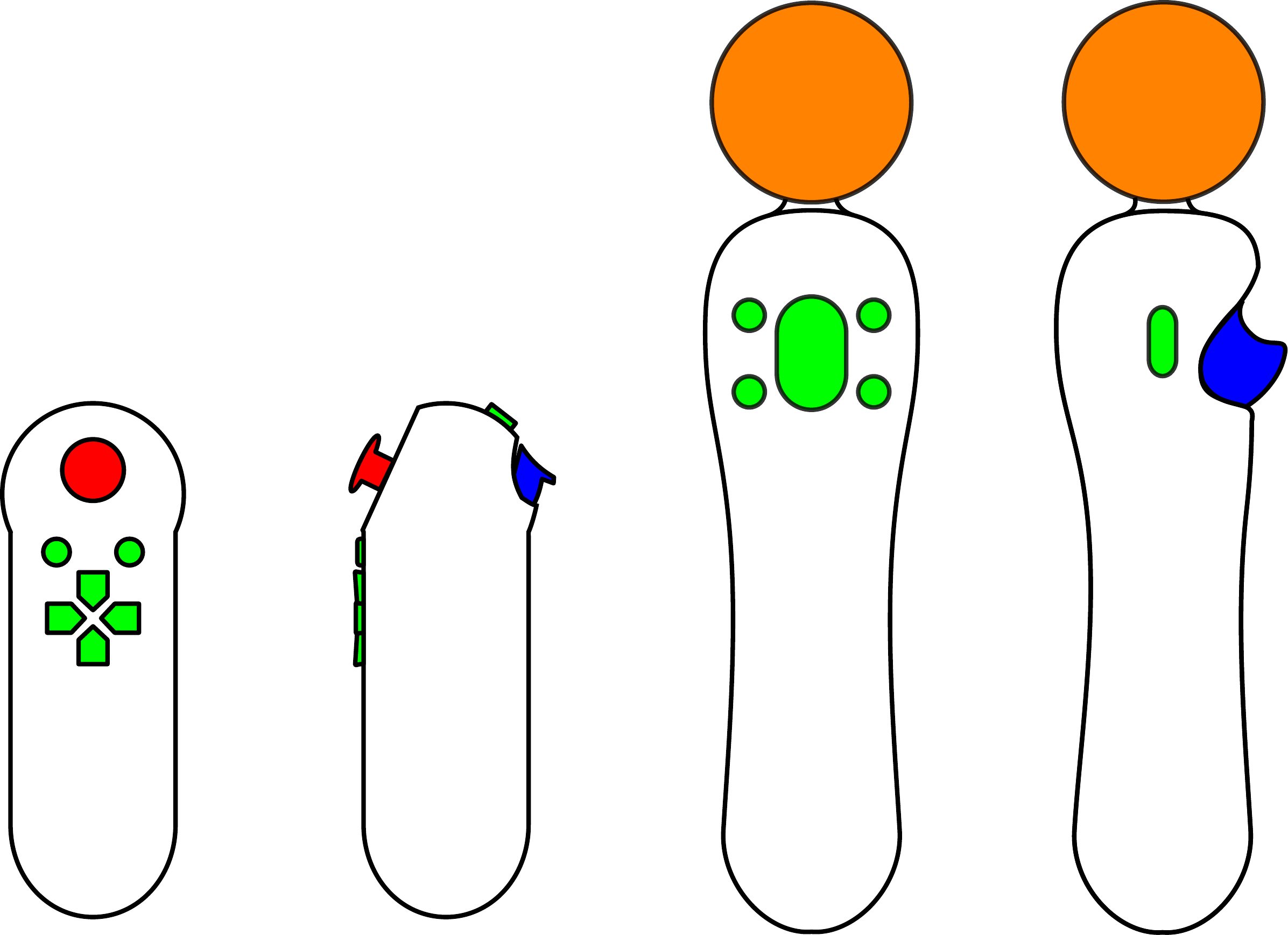}
	\caption{The Sony PlayStation Move Nav-Pad (left) and controller (right). 
        Both devices provide digital buttons (green) and an analogue trigger
        (blue). The Nav-Pad provides an additional analogue stick (red), while
        the controller has a bulb (orange) on its top that can glow in different colors.}
	\label{fig:move}
\end{figure}

A peak $p$ is defined by its center $c$, width $w$, and height $h$:
\begin{equation}
p(x) = \begin{cases}
       h\cdot\sin\left( \frac{\pi}{2w} (x-c+w) \right) & c-w \leq x \leq c+w\\
       0 & \text{otherwise}
       \end{cases}
\end{equation}
(Alternatively, different window functions could be used).

The result of a transfer function $t$ for a voxel value $x$ is then defined as
the result of alpha-blending the peak colors using the peak value $p(x)$ as the
opacity value.

\begin{figure}[t]
	\centering
	\includegraphics[width=.66\linewidth]{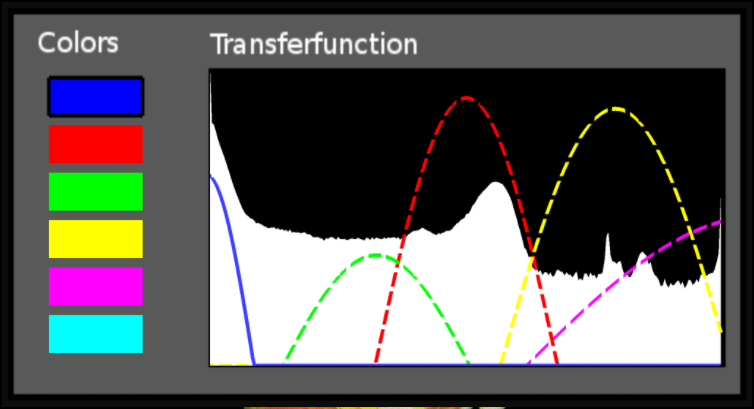}
	\caption{Transfer function defined by $n=5$ peaks. Each peak assigns
        color and opacity information to a voxel value range. The histogram of
        voxel values is displayed in white in the background.}
	\label{transfer_widget}
\end{figure}

In practical use, only a small number $n < 8$ of peaks is required, since more
peaks tend to clutter up the visualization result. An example is given in
Fig.~\ref{transfer_widget}.

This transfer function model significantly reduces the number of parameters that
a user has to modify, while still allowing flexible and powerful transfer
functions.

The user can add a peak to a transfer function and select its color from a list
of predefined colors using digital buttons on the Move Nav-Pad (see
Fig.~\ref{fig:move}). Similarly, peaks can be deleted, temporarily disabled, or
selected for parameter adjustment using additional Nav-Pad buttons.

To change the center $c$, width $w$, and height $h$ of a peak, the Move
controller is used. We tried different combinations of mapping these three peak
parameters to the $x$-, $y$-, and $z$-axes of the Move controller. Using the
$z$-axis proved to be unintuitive and therefore difficult to control. Our
current solution is that the user has to choose (using buttons on the Move
controller) to adjust either $c$ and $h$ or $w$. This has the advantage that both
adjustments take place in the $x/y$-plane. The reason for separating $w$ from
the other parameters was that the visualization result is especially sensitive
to changes of peak widths, so that users tend to first define position and
height of a peak and then fine-tune the result by adjusting its width.

To provide the user with visual feedback about the current transfer function
properties and the selection state, we display an overview widget at a fixed
position in the Virtual Environment, as shown in Fig.~\ref{transfer_widget}.
Note that this widget is for informational purposes only and does not require
traditional point-and-click functionality.

As an additional aid, we set the color of the glowing Move controller bulb
to the color of the transfer function that is currently selected. Experienced
users can use this feedback to determine the current state of transfer function
manipulation without looking at the overview widget.


\subsubsection{Navigation}

Navigation is implemented by performing translation and rotation using the
tracked Move controller. Translation is active while the largest digital
button of the Move controller is pressed, while rotation is active
while the analogue trigger of the Move controller is pressed. See
Fig.~\ref{fig:move}. Translation works by directly applying Move controller
position changes to the volume. Rotation works by mapping horizontal
controller movements to volume rotations around the $y$-axis, vertical
movements to volume rotations around the $x$-axis, and controller rotations
around the $z$-axis to volume rotations around the $z$-axis.

An obvious alternative would be to directly apply both Move controller position
and orientation changes to the volume while in navigation mode, but separating
translation and orientation in the way described above allows a more
fine-grained control of movement, which is useful to examine smaller volume
areas in detail. Furthermore, mapping controller movements to rotations instead
of directly using the controller orientation avoids uncomfortable wrist positions.
Requiring a button to be pressed for navigation mode allows the
user to move freely in the Virtual Reality environment without unintentionally
moving the volume.

\subsection{Implementation}

The physical setup is described by Fig.~\ref{vrlab_constraints}.

Our software implementation is based on the Equalizer framework for parallel and
distributed rendering~\cite{eilemann09equalizer}. This allows the application to
run across the six nodes of our render cluster, each of which renders one of the
stereo channels using two graphics cards.

We used the official Move.Me SDK from Sony for connecting the PlayStation Move
system to Equalizer. The controller sends its sensor data via Bluetooth to the
PlayStation console, on which the Move.Me SDK runs as a server application. Our
application acts as a client to this server and receives position, orientation,
and button state data via network UDP packets. This data is transformed to
custom input events and handed over to the Equalizer event handling mechanisms
to allow consistent input event handling.

The GPU-based volume ray caster is based on a ray casting shader implementation
provided by the Visualization Library
project\footnote{\url{http://www.visualizationlibrary.org}}. The ray caster is
straightforward but proved sufficient for our purposes while being fast enough
for interactive use in a Virtual Reality environment.

\begin{figure*}[t]
\centering
\minipage[t]{0.32\linewidth}
	\includegraphics[width=\linewidth]{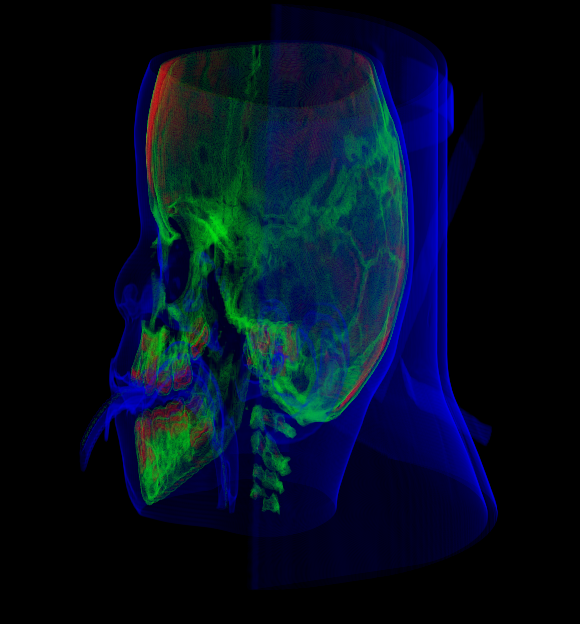}
	\includegraphics[width=\linewidth]{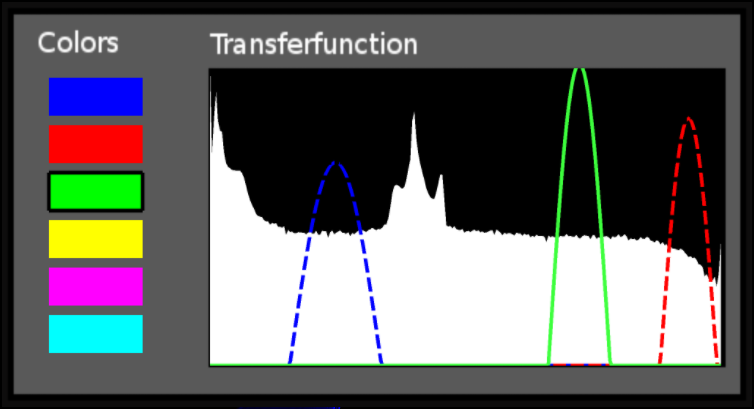}
	\caption{Predefined result}
	\label{experiment_screenshot_reference}
\endminipage\hfill
\minipage[t]{0.32\linewidth}
	\includegraphics[width=\linewidth]{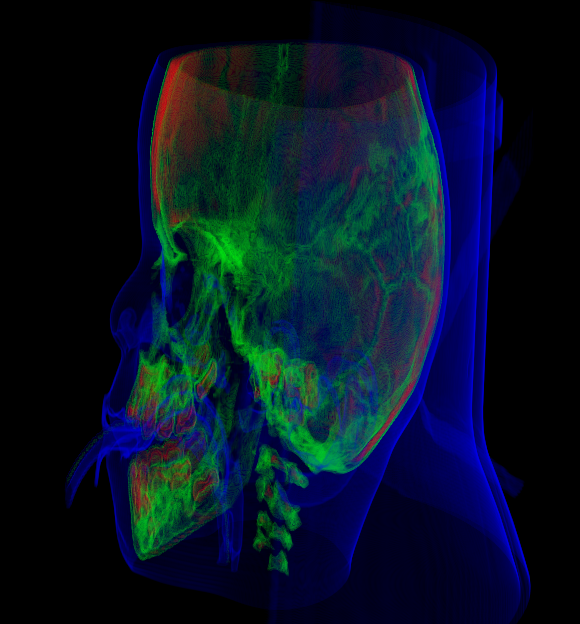}
	\includegraphics[width=\linewidth]{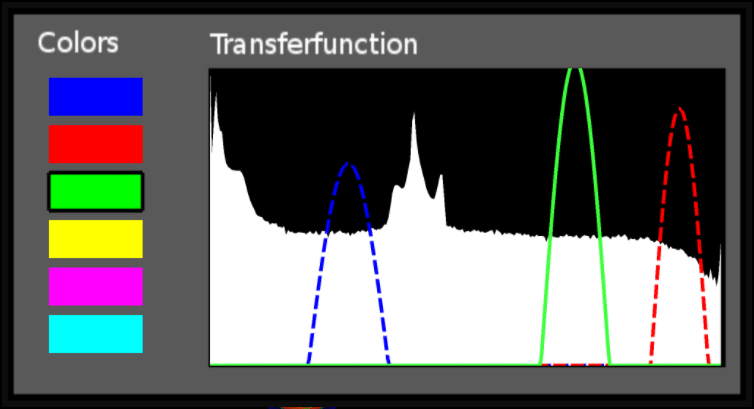}
	\caption{Result of the experienced user}
	\label{experiment_screenshot_experienced}
\endminipage\hfill
\minipage[t]{0.32\linewidth}
	\includegraphics[width=\linewidth]{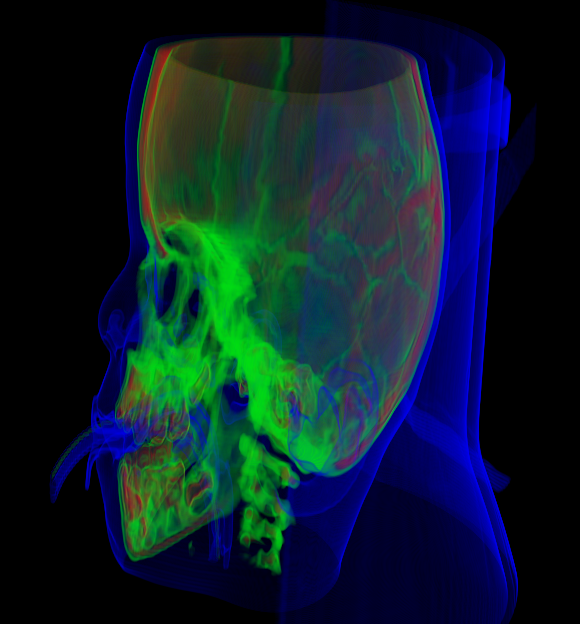}
	\includegraphics[width=\linewidth]{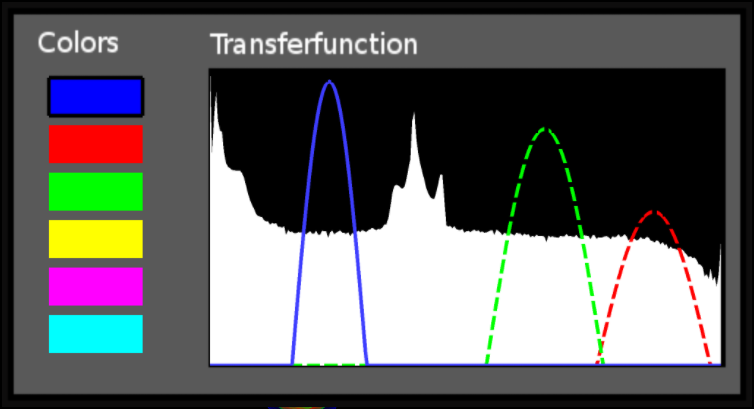}
	\caption{Result of the inexperienced user}
	\label{experiment_screenshot_inexperienced}
\endminipage
\end{figure*}

\section{Initial Results}
\label{sec:results}

In the figures throughout this paper, we used the ``Baby Head'' data set
available from the volume library of Stefan
Roettger\footnote{\url{http://schorsch.efi.fh-nuernberg.de/data/volume/}}.

As a first test of our concept, we performed an experiment involving 
one test user with experience in both Virtual Reality and Volume Rendering
applications and another test user with no experiences in these domains.

The task for these two test users was to reproduce a visualization result for
the ``Baby Head'' data set, starting from a default transfer function. The
predefined visualization result was produced with a transfer function consisting of three
peaks that separate the skeleton (green), the teeth (red), and the skin (blue).
See Fig.~\ref{experiment_screenshot_reference}.

The experienced test user was able to solve the task in less than half a minute
with good results, shown in Fig.~\ref{experiment_screenshot_experienced}.
After an introduction to controller usage and button configuration, the
inexperienced user was able to achieve a satisfying result, shown in
Fig.~\ref{experiment_screenshot_inexperienced}, in approximately one minute.

In this and similar experiments with several test data sets (only one of
which is shown here), the simplified transfer function model was powerful enough
and the transfer function manipulation was precise enough to highlight
interesting aspects the data. More demanding data sets might require more
fine-grained control, which may require changes to the transfer function model
and/or adjustments to the interface sensitivity.

The two-handed interface requires some coordination between both hands which
inexperienced users are unaccustomed to. However, after some practice, this
does not seem pose a problem.

Another challenge especially for inexperienced users is to remember the mapping
between controller movements and buttons to the various interaction
functionalities of the application. We plan to integrate an online help function
that displays images of the controllers on screen, along with schematic
descriptions of their usage, in order to avoid the need for leaving the Virtual
Reality environment to look up user interface documentation.

Despite these shortcomings, we believe the approach has the potential to be
usable in real-world Volume Rendering applications.

\section{Conclusion}
\label{sec:conclusion}

We propose a 3D user interface concept for Volume Rendering in Virtual
Environments. Unlike previous approaches, this concept does not try to map
traditional 2D point-and-click concepts to the Virtual Environment; instead, it
is based on a set of intuitive user actions using the Sony PlayStation Move
controller system.

For this purpose, a simplified transfer function model was designed that allows
a reduction of interaction complexity. This comes at the cost of reduced
flexibility and precision when compared to a traditional 2D desktop interface,
but we believe that our system is still flexible and precise enough for
exploration of most volume data sets, while allowing faster and more intuitive
manipulation of transfer functions.

In the future, we would like to refine the interface based on user feedback.
Furthermore, it would be interesting to explore the possibility to extent our
approach to two-dimensional transfer functions.


\bibliographystyle{abbrv}
\bibliography{paper}
\end{document}